\documentclass[runningheads,a4paper]{llncs}

\usepackage{amssymb}
\usepackage{graphicx}

\usepackage{amsmath}
\usepackage{amsmath,bm}

\usepackage{amssymb}
\usepackage[T1]{fontenc}
\usepackage[latin9]{inputenc}
\usepackage{array}
\usepackage{float}
\usepackage{amstext}
\usepackage{graphicx}
\usepackage{float}

\makeatletter

\providecommand{\tabularnewline}{\\}
\floatstyle{ruled}
\newfloat{algorithm}{tbp}{loa}
\providecommand{\algorithmname}{Algorithm}
\floatname{algorithm}{\protect\algorithmname}

\usepackage{enumitem}
\setlist{nolistsep}

\@ifundefined{showcaptionsetup}{}{%
 \PassOptionsToPackage{caption=false}{subfig}}
\usepackage{subfig}
\makeatother

\title{Classification of message spreading in a heterogeneous social network}

\author{Siwar Jendoubi\inst{1,2}
\thanks{These research works and innovation are carried out within the framework of the device MOBIDOC financed by the European Union under the PASRI program and administrated by the ANPR.}
 \and Arnaud Martin\inst{2} \and \\ Ludovic Li\'etard\inst{2}
 \and Boutheina Ben Yaghlane\inst{1}}
\authorrunning{Siwar Jendoubi et al.} 

\institute{LARODEC, University of Tunis, Avenue de la libert\'e, 2000 Le Bardo, Tunisie
\and
IRISA, University of Rennes 1, Rue E. Branly, 22300 Lannion, France
\\
jendoubi.siwar@yahoo.fr,  Arnaud.Martin@univ-rennes1.fr, ludovic.lietard@univ-rennes1.fr, boutheina.yaghlane@ihec.rnu.tn}

\begin{document}

\maketitle

\begin{abstract}
Nowadays, social networks such as Twitter, Facebook and LinkedIn become
increasingly popular. In fact, they introduced new habits, new ways
of communication and they collect every day several information that
have different sources. Most existing research works focus on the
analysis of homogeneous social networks, {\em i.e.} we have a single type
of node and link in the network. However, in the real world, social
networks offer several types of nodes and links. Hence, with a view
to preserve as much information as possible, it is important to consider social
networks as heterogeneous and uncertain. The goal of our paper is
to classify the social message based on its spreading in the network
and the theory of belief functions. The proposed classifier interprets
the spread of messages on the network, crossed paths and types of
links. We tested our classifier on a real word network that we collected
from Twitter, and our experiments show the performance of our belief
classifier.
\end{abstract}

\section{Introduction}

Nowadays, social networks such as Twitter, Facebook and LinkedIn become
increasingly popular. In fact, they introduced new habits and new
ways of communication. Besides, one of the distinguishing features of on-line
social networks is the information spreading through social links.
This is due to the ``word-of mouth'' exchanges, {\em i.e.} user-to-user
exchanges, which makes the information more accessible and it spreads
and reaches a large scale in few minutes. The volume and the dynamic
of the exchange has attracted the attention of research communities.
This research is motivated by the fact that the study of the diffusion
of information is useful for understanding the dynamic behind social
networks and the evolution of human relationships. Thus, they have
focused on the processing of such data to extract high
quality information, this information may be an important event, it
can also be useful for optimizing business performance, or even for
preventing terrorist attacks, etc.

The processing of a social network, always, starts by studying its structural
properties, in fact the simple visualization of the network cannot
give us a clear analysis about it. In the literature, we found a lot
of structural properties measures like the degree, the betweenness,
the closeness, the eigenvector centrality, etc. Quantifying structural
properties and interpreting them will be essential to characterize
the behavior of social actors, their position in the network, their
interactions and how do they diffuse the information. Hence, the analysis of the network structural properties is an essential step when we study and model information propagation.

In our work we are interested in the classification of the spreading
of the information in a heterogeneous social network. We assume that each type of content has some specific behavior when it propagates in the network. Hence, we propose a new algorithm of information propagation
in a heterogeneous social network that takes into account the behavior
of the content to be propagated. Therefore, we introduce an evidential algorithm
to classify the propagation of the information through the network.

In the next section, we outline the literature review of the information
propagation in social networks, the social message classification
and the theory of belief functions. In section three, we introduce our algorithm
of information propagation in a heterogeneous social network. In section
four, we present our classification algorithm. Finally, we present
our experiments in the fifth section.

\section{Literature review}

\subsection{Information propagation in social networks}

Information dissemination is a wide research domain that attracted
the attention of researchers from various field such as physics and
biology. We find the family of epidemiological models that are used
to understand how diseases spread through populations. The simplest
version is SI (\textit{Suspected-Infected}), in this model, an individual
is suspected if he has not the disease yet but he can catch it and
become infected. This model was extended and many other version appeared
to model specific diseases. Hence, we find SIS model (\textit{Suspected-Infected-Suspected}),
SIR model (\textit{Suspected-Infected-Recovered}), SIRS model (\textit{Suspected-Infected-Recovered-Suspected}),
etc. The reader can refer to \cite{Anderson91,Newman10} for further
details.

Computer scientists are generally interested in studying information
propagation in on-line social networks. Mainly, their goal is to
develop a model that simulates the diffusion process. Basic models
are \textit{Linear Threshold Model} (LTM) \cite{Granovetter78} and
\textit{Independent Cascade Model } (ICM) \cite{Goldenberg01}. They
assume the existence of a structure of a directed graph where each
node can be activated or not knowing that you can not inactivate already
activated nodes. The ICM model requires a probability distribution
which must be associated with each link and LTM requires a degree
of influence that must be set on each link and a threshold of influence
for each node \cite{Kempe03}. These two models were reused and improved
in a lot of works like \cite{Galuba10,Saito11}.

In this paper, we focus on information propagation in a heterogeneous
social network, {\em i.e.} on which we find several types of links and/or nodes. 
In fact, in real word social networks we find many types
of objects (users, groups, applications, etc) that are connected {\em via}
many types of social links (friendship, membership, colleague, etc). 
Information dissemination in homogeneous social networks has been
widely studied and the reader can refer to \cite{Gille13} for a
recent survey. Now, research works start focusing on the processing
of heterogeneous social networks. We find the work of \cite{Sermpezis13}
that simulates the propagation of the information in heterogeneous
social networks based on the configuration model approach. In \cite{Li12}, authors 
propose to consider the behavior of individuals to model the influence
propagation, their model is based on a heterogeneous social network.

\subsection{Social message classification}

Social message classification approaches, presented in the literature, are generally based
on the content of the information and text mining techniques. They search to classify the user generated content to positive or negative about a some specific product. This task is so called sentiment classification and it is used to mine opinions. It starts by an item and/or feature extraction step, then it compares the extracted items and/or features to an existing corpus, finally comes the sentiment classification that can be based on items, features or both of them \cite{Lo09}. We find the work of \cite{Mostafa13} in which the author
used a random sample of 3516 tweets to classify the feelings of consumers
with respect to well-known brands. He classified the opinions (tweets)
into positive and negative to see what is the most dominant opinion. 
In \cite{He13}, a detailed case study that applies text mining
to analyze unstructured textual content published on Twitter and Facebook
and that talks about three chains of pizza. The reader can refer to
the work of \cite{Nettleton13} for a recent study of the state of
the art of social networks data mining.

\subsection{Theory of belief functions}

\textit{Upper and Lower probabilities} \cite{Dempster67a} was the
first ancestor of the theory of belief functions. Then comes the \textit{Mathematical
theory of evidence} \cite{Shafer76} which defines the basic framework
of information management and processing in the evidence theory, often
called \textit{Shafer model}. The main purpose of the theory of belief functions
is to achieve more reliable, precise and coherent information. Here we present a short introduction of this theory, for more details the reader can refer to \cite{Shafer76}.

Let $\Omega=\left\{ \omega_{1},\omega_{2},\ldots,\omega_{n}\right\} $ be a set of all
possible decisions that can be made in a particular problem, it is
called frame of discernment. The basic belief assignment
(BBA), $m^{\Omega}$, represents the agent belief on $\Omega$, and
it must respect $\sum_{A\subseteq\Omega}m^{\Omega}\left(A\right)=1$.
In the case where we have $m^{\Omega}(A)>0$, $A$ is called focal
set of $m^{\Omega}$. The basic belief assignment can be converted
into other functions defined from $2^{\Omega}$ to $\left[0,1\right]$.
This theory presents a rich framework for information fusion
and combining pieces of information (evidence). 
We find the Dempster's rule \cite{Dempster67a}, the conjunctive and disjunctive combination rule 
\cite{Smets93a}, etc.

\section{Propagation algorithm in a heterogeneous social network}

In this section we introduce an algorithm of information propagation
in a heterogeneous social network. This new algorithm takes four different
inputs which are the number of iterations (stopping condition), the
source of the message, the propagation strategy and the heterogeneous
social network. As output, we have the propagation network that preserves
the traversed paths. Algorithm \ref{alg:Information-propagation-algorith}
shows outlines of our propagation process. It starts by
the source node. First, we verify if the current node is ready (wants) to
propagate the message. Then, for each type of link in the network
we compute the number of neighbors that will receive the message.

\begin{algorithm}
\textbf{Inputs:}
\begin{itemize}
\item \textbf{N:} number of iteration
\item \textbf{S:} source of the message
\item \textbf{Str:} propagation strategy
\item \textbf{Network:} the heterogeneous social network
\end{itemize}
\textbf{Output:}
\begin{itemize}
\item \textbf{PrNet:} propagation network
\end{itemize}
\textbf{Algorithm:}
\begin{enumerate}
\item ReadyNodes.add$\left(S\right)$;
\item \textbf{For} $i=1$ \textbf{to} $N$ \textbf{do}

	\begin{enumerate}
		\item \textbf{for} $j=1$ \textbf{to} ReadyNodes.size$\left(\right)$ \textbf{do}

		\begin{enumerate}
			\item Node $\leftarrow$ReadyNodes.get$\left(j\right)$;
			\item if$\left(\textrm{Node.propagate\ensuremath{\left(\right)}=True}\right)$
		\end{enumerate}

\hspace{1.2cm}\textbf{foreach} LinkType \textbf{do}

\hspace{2.5cm}$x \leftarrow \textrm{Node.outdegree\ensuremath{\left(\right)}}*\textrm{Node.propagationTendancy\ensuremath{\left(\right)}}$

\hspace{3.5cm}$*\textrm{Str.LinkTypeProportion\ensuremath{\left(\right)}}$;


\hspace{2.5cm}R$\leftarrow \left(\textrm{Node.randomSelection\ensuremath{\left(x,LinkType\right)}}\right)$;
\item Pr.refine$\left(R\right)$;
\item R1.addAll$\left(R\right)$;
\item ReadyNodes.addAll$\left(R1\right)$;
\item R1.clear;
\end{enumerate}
\end{enumerate}
\caption{Information propagation algorithm\label{alg:Information-propagation-algorith}}

\end{algorithm}

We assume that each type of message has some special characteristics
of propagation in the network that is related to the types of links,
so we define a propagation strategy for each type of message. Moreover, 
we consider the tendency of a particular node to
propagate the message as a propagation parameter. Indeed, this parameter
models the fact that a node can choose to distribute  the message to a subset of its contacts (that he selects) or to retain it. The novelty of this 
algorithm is that we consider the type of the message while propagating
it. Moreover our algorithm works with heterogeneous social networks where
we have different types of links.

\section{Classification of information propagation}

The main purpose of this paper is to classify the spreading of the information
through the network in order to characterize its content. In this
section, we introduce our classification process that is composed of
two steps; parameter learning step and the classification step. As
mentioned in the algorithm \ref{alg:Parameters-learning-algorithm},
to learn the parameters of the model we need a set of propagation
networks. First of all, we compute the number of nodes that have received
the message {\em via} each type of link. We do this computation for each
propagation level, {\em i.e.} we call propagation level the number of links
between the source of the message and the target node. Second,
we calculate the accrued effective by summing the effective of each
level with the effective of the one before, this computation is done in order to preserve the propagation history at each propagation level. After that we transform
the effective set of each level to a probability distribution
defined on types of links, this transformation is done for two reasons; the first one, we need a probability distribution for the probabilistic classifier and the second one, it is an essential step to get the basic belief assignment distribution. Finally we transform each probability
distribution to a BBA distribution using the consonant
transformation \cite{Aregui07b,Aregui08a}.

\begin{algorithm}[H]
\textbf{Input:}
\begin{itemize}
\item \textbf{PrNetSet: }a set of propagation networks
\end{itemize}
\textbf{Output:}
\begin{itemize}
\item \textbf{ProbaSet:} a set of probabilities distributions (a probability
distribution by propagation level).
\item \textbf{BbaSet:} a set of BBA distributions (a BBA distribution by
propagation level).
\end{itemize}
\textbf{Algorithm:}

//effective computation

\hspace{-0.15cm}\textbf{Foreach} PrNet \textbf{in} PrNetSet \textbf{do}
\begin{enumerate}
\item \textbf{Foreach} Level \textbf{in} PrNet \textbf{do}

\begin{enumerate}
\item \textbf{Foreach} TypeLink \textbf{do}

\hspace{1.2cm}$N\left(\textrm{TypeLink, Level}\right)\leftarrow N\left(\textrm{TypeLink, Level}\right)$

\hspace{1.2cm}$+\textrm{ComputeNodes\ensuremath{\left(TypeLink\right)}}$;

\end{enumerate}
\end{enumerate}
//Accrued effective calculation

\hspace{-0.15cm}\textbf{For} Level$=2$ \textbf{to} NbrLevels \textbf{do}
\begin{enumerate}
\item \textbf{Foreach} TypeLink \textbf{do}

\begin{enumerate}
\item $N\left(\textrm{TypeLink, Level}\right)\leftarrow N\left(\textrm{TypeLink, Level}\right)$

$+N\left(\textrm{TypeLink, Level}-1\right)$;

\end{enumerate}
\end{enumerate}
//ProbaSet and BbaSet computation

ProbaSet$\leftarrow$ProbabilitiesCalculation$\left(N\right)$;

BbaSet$\leftarrow$ConsonantTransformation$\left(\textrm{ProbaSet}\right)$;

\caption{Parameter learning algorithm\label{alg:Parameters-learning-algorithm}}

\end{algorithm}

Once model's parameters are learned, we can use it to classify new
coming message (propagation network of the message) as shown in
algorithm \ref{alg:Classification-algorithm}. Our classification
algorithm starts by applying the same parameter learning process (algorithm
\ref{alg:Parameters-learning-algorithm}) on the propagation network
to be classified. Then for each level in the network we compute the
distance between its probability distribution and the probability distribution
of each propagation strategy, then we choose the class of the nearest 
propagation strategy (with the shortest distance) to be the class of the message in the current
level. The same process is done with BBA distributions as mentioned
in the algorithm.

\begin{algorithm}[H]
\textbf{Input:}
\begin{itemize}
\item \textbf{ProbaSets:} a set of probabilities distributions for each
strategy of propagation.
\item \textbf{BbaSets:} a set of BBA distributions for each strategy of
propagation.
\item \textbf{PrNet:} The propagation network to be classified
\end{itemize}
\textbf{Output:}
\begin{itemize}
\item In order to see the impact of the level of propagation on the classification
results, in our output we have a class by level.
\end{itemize}
\textbf{Algorithm:}
\begin{enumerate}
\item $\left(\textrm{ProbaPr, BbaPr}\right)\leftarrow\textrm{ParameterLearning}\left(PrNet\right)$;
\item \textbf{For} $i=1$ \textbf{to} NbrStrategies \textbf{do}

\begin{enumerate}
\item \textbf{Foreach} Level \textbf{do}

\begin{enumerate}
\item ProbaDist$\left(i,Level\right)$ $\leftarrow\textrm{Distance}\left(\textrm{ProbaPr, ProbaSets}\left(i\right)\right)$;
\item BbaDist$\left(i,Level\right)$ $\leftarrow\textrm{Distance}\left(\textrm{BbaPr, BbaSets}\left(i\right)\right)$;
\end{enumerate}
\end{enumerate}
\item \textbf{Foreach} Level \textbf{do}

\begin{enumerate}
\item ProbaClasses$\left(Level\right)\leftarrow\textrm{StrategyMinDistance}\left(\textrm{ProbaDist}\left(:,Level\right)\right)$;
\item BbaClasses$\left(Level\right)\leftarrow\textrm{StrategyMinDistance}\left(\textrm{BbaDist}\left(:,Level\right)\right)$;
\end{enumerate}
\end{enumerate}
\caption{Classification algorithm\label{alg:Classification-algorithm}}

\end{algorithm}

\section{Experiments and results}

In this section, we present some experiments to show the power of the proposed
evidential classification algorithm. 

\subsection{Data description}

\begin{figure}
\begin{centering}
\includegraphics{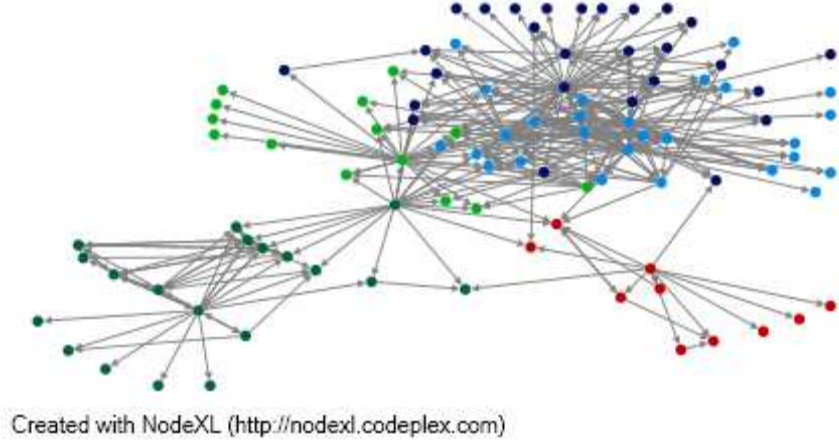}
\par\end{centering}

\caption{Network visualization\label{fig:Network-visualization}}

\end{figure}

We used NodeXL V 1.0.1.245 \cite{Hansen11} to collect social network
data from Twitter. We collected the network shown in figure \ref{fig:Network-visualization}. 
It is a directed network in which nodes are Twitter users. 
Table \ref{tab:Data-characteristics}
shows the characteristics of our network data.

\begin{table}[h]
\caption{Data characteristics\label{tab:Data-characteristics}}

\begin{centering}
\begin{tabular}{|c|c|>{\centering}p{20mm}|c|c|c|}
\hline 
{\small{Vertices }} & {\small{Edges }} & {\small{Geodesics distance}} & {\small{Betweenness}} & {\small{Closeness}} & {\small{Eigenvector}}\tabularnewline
\hline 
{\small{97}} & {\small{350}} & {\small{6}} & {\small{184.99}} & {\small{0.004}} & {\small{0.01}}\tabularnewline
\hline 
\end{tabular}
\par\end{centering}

\end{table}

As mentioned above, we need a heterogeneous social network to test proposed algorithms.
Therefore, we used the structure of the network collected
from Twitter and we generated, randomly, the types of links. We assume
four types of link in the network which are ``Professional'',
``Familial'', ``Friendly'' and ``Undefined''. Then we obtained
a heterogeneous social network that is used as input for our propagation
algorithm.

\subsection{Experiment configuration}

In the following experiments, we defined three different propagation strategies 
for three types of messages which are: ``Spam'', ``Professional''
and ``Familial''. Each strategy is defined as the proportion of
the nodes that will receive the message from each type of links. Hence,
we have to define four proportions for each propagation strategy.
To be as near as possible to the reality, we added a noise rate to
the strategy. We note that the noise value can be added or removed
from the proportions of kind of messages.  We used the euclidean distance for the probabilistic
classifier:

\begin{equation}
d_{E}\left(Pr_1,Pr_2\right)=\sqrt{\sum_{i=1}^{card}\left(Pr_1\left(i\right)-Pr_2\left(i\right)\right)^{2}}\label{eq:distance}
\end{equation}
and the Jousselme distance \cite{Jousselme01a} for the evidential
one:
\begin{equation}
d_{J}\left(m_{1},m_{2}\right)=\sqrt{\frac{1}{2}\left(m_{1}-m_{2}\right)^{T}\underset{=}{D}\left(m_{1}-m_{2}\right)}
\end{equation}
such that $\underset{=}{D}$ is an $2^{n}\times2^{n}$ matrix and
$D\left(A,B\right)=\frac{|A\cap B|}{A\cup B}$. We fixed the number of levels in the network
to three (three iterations in the propagation algorithm). Then we
run the proposed propagation algorithm to create a training set for each propagation
strategy, we fixed the size of the strategy training set to 100 propagation
networks. Also, we created a testing set of size 100.

\subsection{Results and discussion}

\begin{figure}
\begin{centering}
\subfloat[Probabilistic results\label{fig:Probabilistic-results}]{\includegraphics[scale=0.45]{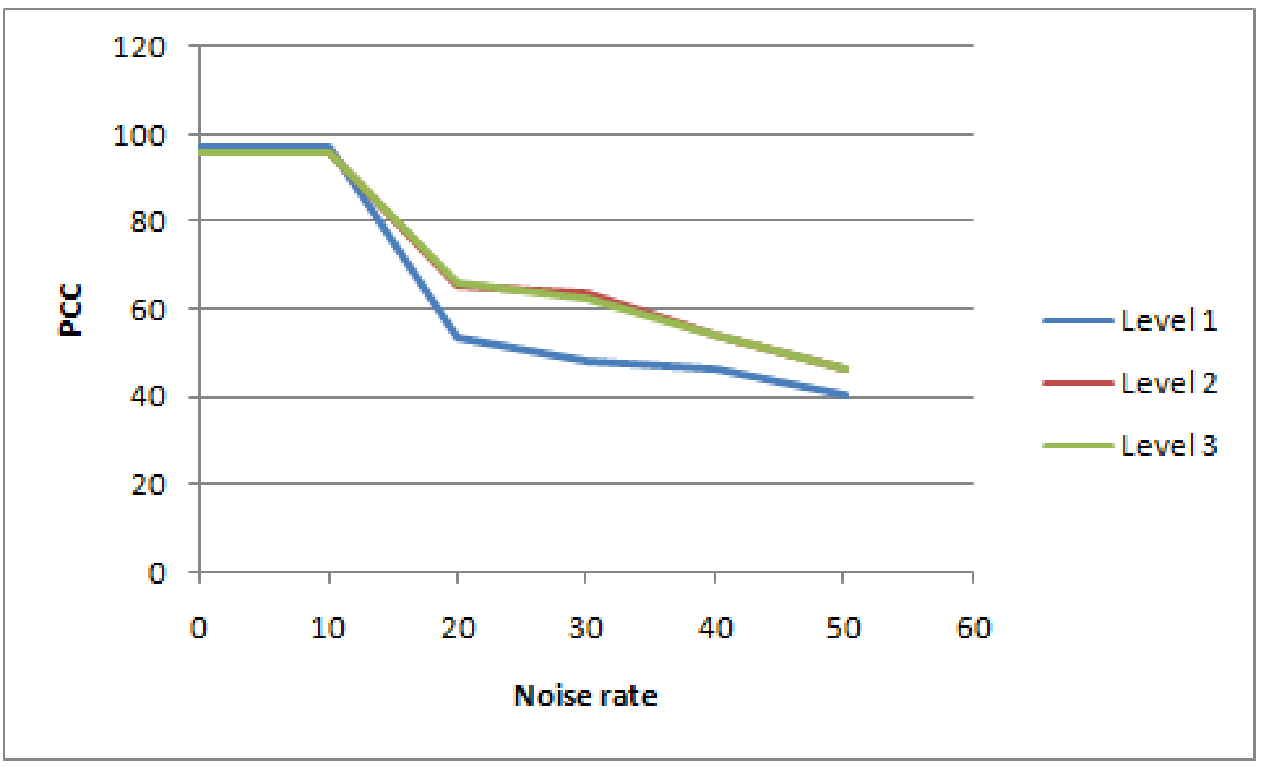}} 
\subfloat[Evidential results\label{fig:Credal-results}]{\includegraphics[scale=0.5]{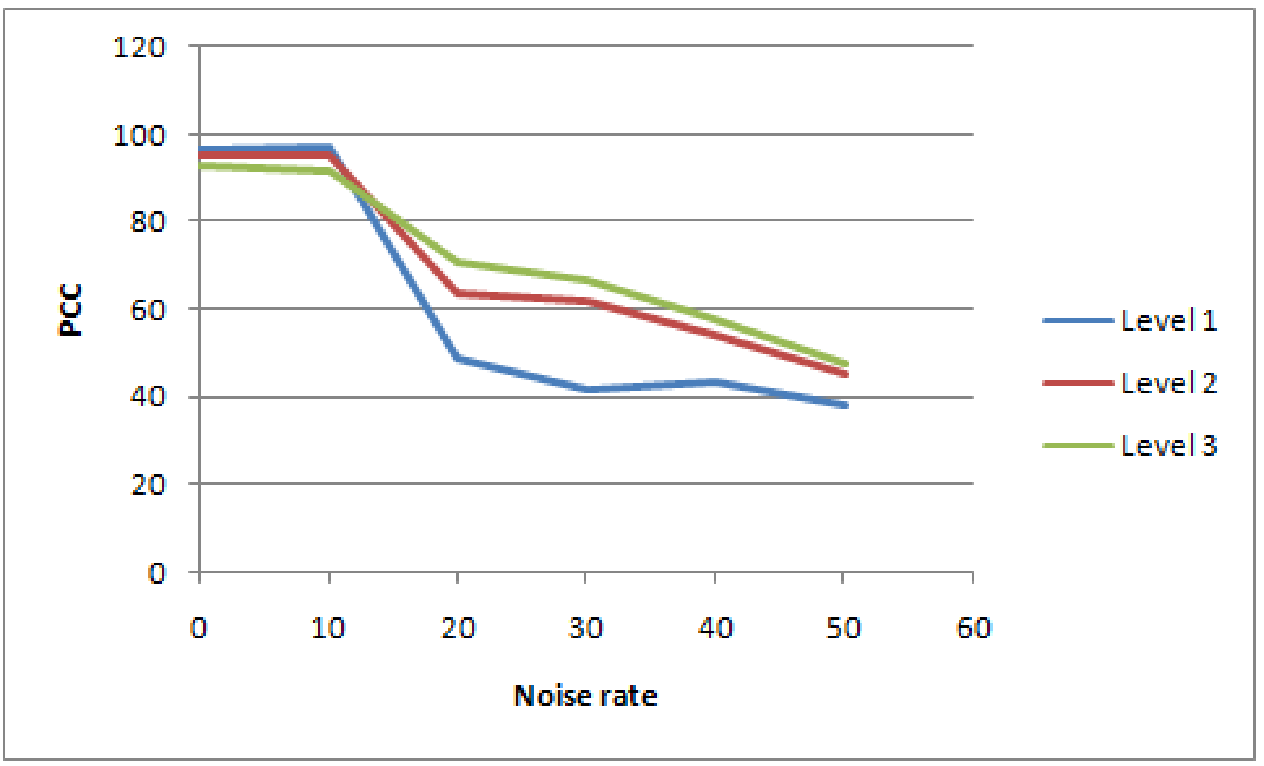}}
\par\end{centering}

\caption{The impact of the propagation level on the PCC\label{fig:LevelImpact}}

\end{figure}

In this section we present our results and a comparison between the probabilistic and the
evidential classifier. To obtain
accurate results we turned the experimental process ten times and we take the
mean of the percentage of correctly classified (PCC) propagation networks.
Figure \ref{fig:LevelImpact} shows the impact of the propagation
level on the PCC of the probabilistic results (figure \ref{fig:Probabilistic-results})
and the evidential results (figure \ref{fig:Credal-results}). Figures
\ref{fig:Probabilistic-results} and \ref{fig:Credal-results} illustrate that
the PCC increases when the propagation level increases and
we observe this fact starting from the noise level 20\%. In figure
\ref{fig:Probabilistic-results} we observe that the curve of the second level
 coincides with the curve of third level and practically there
is no improvement in the PCC. However, in figure \ref{fig:Credal-results}
 (evidential results), we note that the PCC increases with
the propagation level, this fact is observed starting from the noise rate 20\%. Hence, we
have the PCC of the third level greater than the PCC of the first and the second levels, and the PCC of the second level is higher than the PCC of
the first one. Therefore, more the message propagates in the network, more we can characterize it. 

\begin{figure}
\begin{centering}
\includegraphics[scale=0.6]{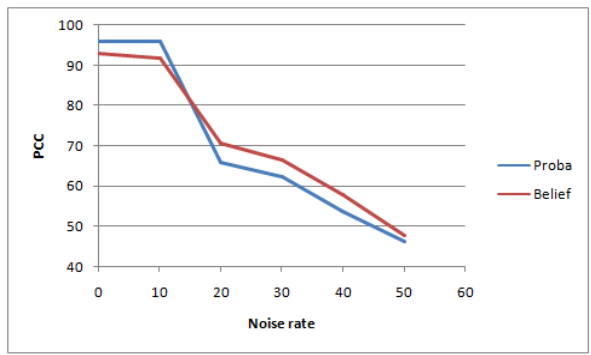}
\par\end{centering}

\caption{Comparison between probabilistic results and evidential results (level
three)\label{fig:Comparision-Belief-proba}}

\end{figure}

In figure \ref{fig:Comparision-Belief-proba}, we compare
the probabilistic and the evidential results of the third propagation
level. We note that without noise (0\%) the probabilistic PCC is
about 96\% (with a 95\% confidence interval of $\pm1.27$) and the
evidential PCC is equal to 93\% (with a 95\% confidence interval of
$\pm1.60$), but in real world social networks the absence of the
noise is an ideal fact and cannot be realistic. When the noise rate
increases, the curve shows that the percentage of correctly classified
propagation networks (messages) decreases. However, we see that the
evidential (Belief) PCC starts to be greater than the probabilistic (Proba)
one. We observe this fact from the noise rate 20\% where we have an evidential PCC equals
to 70.7\% ($\pm4.33$) and a probabilistic
PCC equals to 65.8\% ($\pm4.18$).
Thus, we can conclude that the evidential classifier is more robust
against the noise and gives better classification rates than
the probabilistic classifier.

\section{Conclusion}

To conclude, we presented a state of the art of the information propagation,
classification of social messages and the evidence theory.
Then, we proposed an algorithm of information propagation in
a heterogeneous social network. Thereafter we introduced a new evidential
classification approach that classifies message propagation in a heterogeneous
social network. Finally, we presented some experiments and we noticed
the performance of the evidential classifier against the probabilistic
one in noisy cases. Moreover, we observed that when the propagation
level increases, the message class becomes more accurate and more realistic. 

For future works, we will compare our propagation algorithm
with previous algorithms. Also, we will search to improve it by the
management of the uncertainty and the imprecision related to types
of relationships between social actors. Our next goal is therefore
to define a message propagation algorithm that takes into account
the uncertainty of the types of relationships that is defined on the
links, also we will search to consider the heterogeneity of nodes in the network. Second, we will run our classification algorithm with a more
complex heterogeneous social network in order to prove its applicability. 

\bibliographystyle{splncs03}
\bibliography{biblio}

\end{document}